\documentclass{article}
\usepackage{spconfa4} 
\usepackage{amsmath,graphicx,url,times,amsmath,amssymb,acronym,balance}
\usepackage{color,booktabs,algorithm,balance}
\usepackage[noend]{algpseudocode}
\usepackage[table]{xcolor}
\usepackage{cite}
\usepackage{lipsum}

\title{Multi-label audio classification with a noisy zero-shot teacher}
\name{Sebastian Braun, Hannes Gamper} 
\address{Microsoft Research Redmond, WA, USA\\
\{sebastian.braun, hannes.gamper\}@microsoft.com}
\acrodef{ACD}{Audio content detection}
\acrodef{STFT}{short-time Fourier transform}
\acrodef{SNR}{signal-to-noise ratio}
\acrodef{segSNR}{segmental signal-to-noise ratio}
\acrodef{SRR}{signal-to-reverberation ratio}
\acrodef{PDF}{probability density function}
\acrodef{WER}{word error rate}
\acrodef{SPP}{speech presence probability}
\acrodef{DNN}{deep neural network}
\acrodef{RNN}{recurrent neural network}
\acrodef{CNN}{convolutional neural network}
\acrodef{FC}{fully connected}
\acrodef{CRN}{convolutional recurrent network}
\acrodef{LSTM}{long-term short-term}
\acrodef{GRU}{gated recurrent unit}
\acrodef{FF}{feed forward}
\acrodef{ReLU}{rectified linear unit}
\acrodef{DL}{deep learning}
\acrodef{MAC}{multiply-accumulate}
\acrodef{CRUSE}{Convolutional Recurrent U-net for Speech Enhancement}
\acrodef{FD}{frequency-domain}
\acrodef{DNS}{deep noise suppression}
\acrodef{SED}{sound event detection}
\acrodef{ASR}{automatic speech recognition}
\acrodef{ERB}{equivalent rectangular bandwidth}
\acrodef{RSNR}{reverberant speech-to-noise ratio}
\acrodef{CLAP}{Contrastive Audio Language Pretraining}
\acrodef{LLM}{Large Language Model}
\acrodef{FN}{false negatives}
\acrodef{FP}{false positives}
\acrodef{AP}{average precision}
\acrodef{AUROC}{area under receiver-operating curve}
\acrodef{BCE}{binary cross-entropy}

\definecolor{matlab1}{rgb}{0, 0.4470, 0.7410}
\definecolor{matlab2}{rgb}{0.8500, 0.3250, 0.0980} 
\definecolor{matlab3}{rgb}{0.9290, 0.6940, 0.1250} 
\definecolor{matlab4}{rgb}{0.4940, 0.1840, 0.5560} 
\definecolor{matlab5}{rgb}{0.4660, 0.6740, 0.1880}

\begin{document}
\ninept
\maketitle
\begin{abstract}
We propose a novel training scheme using self-label correction and data augmentation methods designed to deal with noisy labels and improve real-world accuracy on a polyphonic audio content detection task. The augmentation method reduces label noise by mixing multiple audio clips and joining their labels, while being compatible with multiple active labels. We additionally show that performance can be improved by a self-label correction method using the same pretrained model. Finally, we show that it is feasible to use a strong zero-shot model such as CLAP to generate labels for unlabeled data and improve the results using the proposed training and label enhancement methods. The resulting model performs similar to CLAP while being an efficient mobile device friendly architecture and can be quickly adapted to unlabeled sound classes.
%
\end{abstract}
\section{Introduction}
\label{sec:intro}
\ac{ACD} is an important problem for streaming platforms, operating systems and playback devices. The task is similar to audio tagging, i.e., labeling sounds present in a given audio segment of typically several seconds length or longer. In contrast to general audio tagging, \ac{ACD} may consist of a small number of higher level labels or super-classes, e.g.\ speech, music, traffic, machines, animals, etc., where each label can include a multitude of specific sounds. For example, the category \emph{music} includes countless genres and styles; \emph{voice} includes different languages, types (spoken, singing, whispering, emotional variations), demographic attributes etc.
Applications include automatic content sorting or retrieval, informing hearing impaired users, scenarios where no audio playback is possible, awareness augmentation or assistance on wearable devices, and content-adaptive sound tuning on playback devices.

Audio event detection, scene classification, tagging and captioning \cite{Virtanen2018} have been researched thoroughly in the recent years with large success using deep learning. The DCASE research challenges with already 10 editions \cite{Plumbley2023} contributed substantially to the research progress. However, a remaining gap between existing work and practical applications is that most methods assume a single active or dominant class per audio clip or segment, i.e., \emph{monophonic} sound events, which is unrealistic for many real-world \emph{polyphonic} scenarios: Only one of seven tasks in DCASE considers polyphonic events. While there is some work addressing multi-label sound classification \cite{Cakir2017}, many advanced solutions are developed for the single exclusively active label case and cannot be applied easily to the general multi-label problem. 
Another rarely addressed problem is that large-scale datasets usually have noisy labels due to human annotation errors, design flaws, re-purposed data from different tasks, or erroneous machine labels. Therefore, when real-world designing detection systems, noisy labels have to be expected and should be accounted or mitigated to obtain a high confidence prediction system.

The authors in \cite{Phan2022} argue that multi-label fits the reality better but often multi-class, i.e., only one active sound type per clip, seems to perform better due to learning more class interactions, resulting in a cumbersome handling for practical systems. They propose to handle the problem by creating combinatorial polyphonic classes and use a divide and conquer group splitting to keep the exploding number of classes tractable.
The Mixup \cite{Zhang2018} training augmentation method proposed for the multi-class detection in the image domain is often adapted without modification to the audio domain and other tasks. However, a potential shortcoming of this approach is again the inherent assumption of a multi-class classification problem by weakening the binary labels through weighted mixing. The second issue is applying the same linear mixing factor to both data (audio) and label domain. 
In this work, we investigate methods to better model this multi-label problem by proposing a modified multi-label mixup, which decouples the data mixing from the label joining.

\cite{Fonseca2019} states the problem of noisy and missing labels in popular large-scale training sets like Audioset~\cite{Gemmeke2017} and proposes a method to learn from noisy labels for sound event detection by modifying the cost function to a smooth version.
The follow-up work \cite{Fonseca2020,Gong2021} both propose a teacher-student framework to enhance labels. First, a teacher model is used to obtain label predictions for the whole training set. In \cite{Fonseca2020} \ac{FN}, missing labels, are flagged and masked out. It was found to work best to sort the label predictions and mask out the upper percentage (~5\%) of missing labels from the loss.
In contrast, Gong et al.\ \cite{Gong2021} propose to correct the missing labels by setting \ac{FN} to true when the teacher predictions exceed a per-class confidence threshold. 
In this work, we propose a simpler false label detection based on the absolute performance of the teacher per class. We show that distinct treatment by correcting high confidence false labels, and masking out lower confidence false labels can improve the performance of the student model.

The contributions in this work are summarized as follows:
1) We introduce a modified mixup with label-joining instead of weighted mixing, which is more suited to the multi-label classification assuming full presence of polyphonic sound classes.  
2) We propose a novel label correction and masking approach that treats as false detected labels with high confidence different than false labels detected with lower confidence: The first type is corrected while the second type is masked out. In contrast to existing literature addressing only the more prevalent missing label problem, we also investigate generalization to correct or mask \ac{FP} labels. 
3) We show how these approaches can be used to enhance labels from a large zero-shot sound classifier to train a small efficient model and still obtain state-of-the-art performance. This enables the use of zero-shot models to define new sound classes, train on unlabeled data, and still obtain high accuracy with small models on real-world data. 
%


\section{System overview}
\label{sec:method}

\subsection{System design and problem formulation}
The widely used property of classifiers that all class probabilities $p_c$ have to sum to one $\sum_c p_c = 1$, a property imposed by the widely used \emph{Softmax} function, poses a serious limitation to practical in general polyphonic scenarios. While often assumed, typical classification losses such as cross-entropy do not produce probabilities with guaranteed relation to the relative level or activity between the active classes. This inter-class interaction complicates the use for detection tasks, where a simple threshold on the probabilities is used to detect sound presence. In this case the optimal threshold would depend on the number of active classes, which is unknown in advance, and absence of all classes is not in the design. 
We therefore resort to the multi-label problem design, allowing consistent detection of multiple simultaneously present classes with a global threshold.

We address the problem of multi-label audio content detection or tagging: We define $C$ classes of sounds of interest that we want to detect in a given audio sequence of several seconds length given by vector $\mathbf{x}$. Multi-label classification means that we assume multiple classes can be active simultaneously. We want the classifier to indicate the presence of each active class $c\in \{1,\hdots,C\}$ independently with high probability $y_c$.

\subsection{Proposed ACDnet architecture}
\begin{figure}[tb]
	\centering
	\includegraphics[width=\columnwidth,clip,trim=100 180 130 110]{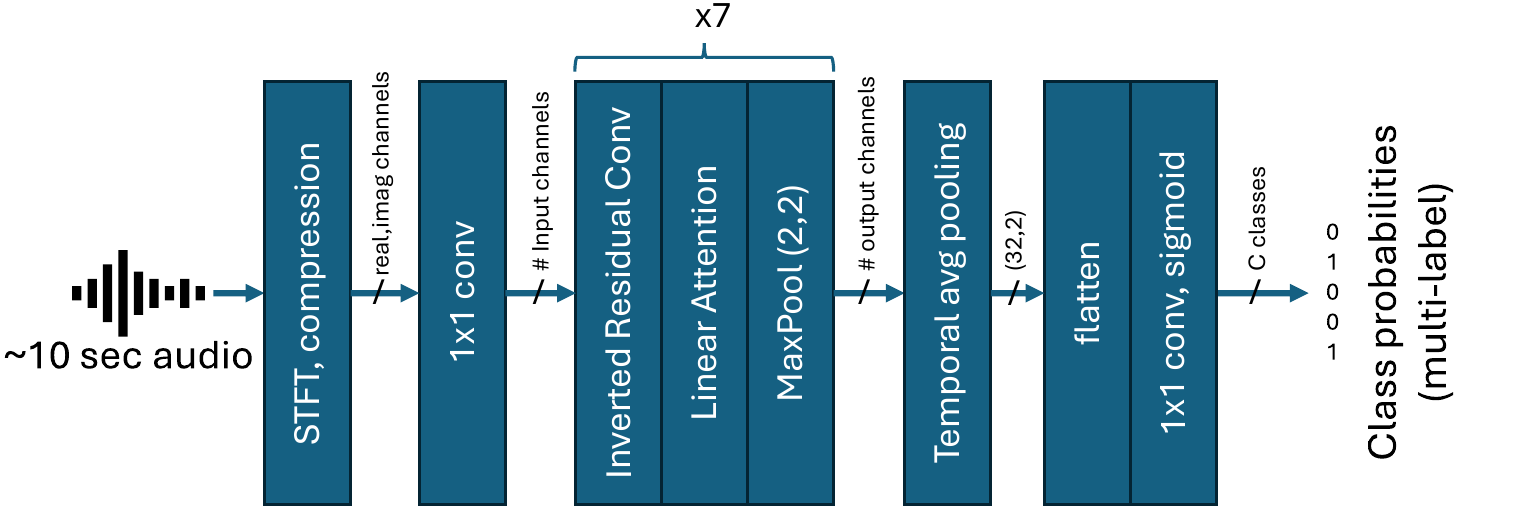}
	\caption{Acoustic content detection architecture}
	\label{fig:ACD_system}
\end{figure}

Our proposed network operates on complex compressed spectral features, obtained from a \ac{STFT} with 50\% overlapping 32~ms windows and applying a $c=0.3$ compression exponent to the magnitudes \cite{Braun2022} . For a 10~s sequence in 16~kHz, this results in a feature map of $624\times257$ time-frequency bins and 2 channels for real and imaginary part as network input.
We design our network as a modified version of the MobileVisionTransformer (MViT) \cite{Mehta2023} using alternating blocks of inverted residual convolution layers \cite{Sandler2018} and attention layers. We utilize linear attention to reduce complexity. After each block we apply $2\times2$ downsampling by max-pooling. Each convolution is followed by batch normalization and parametric ReLU. To increase the receptive field, every other block, we use dilation (2,2) in the convolutional block. We use 7 blocks with filter sizes [32, 64, 128, 256, 128, 64, 32] and $3\times3$ convolution kernels. For sequences of 10~s, this leaves 32 feature maps of size (62,2), where we reduce the temporal dimension by average pooling, resulting in a feature map of $32\times2=64$. This is mapped via a dense layer to the number of classes and a sigmoid activation. The network architecture is shown in Fig.~\ref{fig:ACD_system}.
The resulting network has 248.54K trainable parameters and 1.40G MACs per 10~s of audio including feature extraction.
While our architecture slightly exceeds the constraints in the DCASE challenge Track 1 by a factor of 2 and 4.6 for parameters and MACs, the computational and memory burden is still reasonably low for deployment on consumer and mobile devices. We call this architecture \emph{Acoustic Content Detection network (ACDnet)}.

The network is trained on a \ac{BCE} loss using AdamW optimizer. We use \ac{AP} as the validation metric and define one epoch as 10~k training sequences. The initial learning rate of $1e^{-4}$ is halved after 150 epochs of the validation metric plateauing.

\section{Proposed data augmentation}
\label{sec:data}

We use the 11 classes shown in Tab.~\ref{tab:clap-prompts}. All classes are super-classes encompassing several of the 632 Audioset classes. The grouping of Audioset classes to our super-classes is available as supplementary material material at \url{https://github.com/sebraun-msr/acd_class_mapping}.

\subsection{Data generation}
We use Audioset~\cite{Gemmeke2017}, a large audio dataset with noisy labels similar to what might be found in practice. We train on audio sequences of 10~s length. However, one sequence is sampled with a random start and end time with a minimum length of 5~s. Data from additional audio files is concatenated until the training sequence length is reached. 

Mixup \cite{Zhang2018} is a commonly used technique to increase performance of classifiers by mixing two data samples of index $i,j$ and their labels:
\begin{align}
	\mathbf{x} = \alpha \mathbf{x}_i + (1-\alpha) \mathbf{x}_j \label{eq:origmixup_data}\\
	\mathbf{y} = \alpha \mathbf{y}_i + (1-\alpha) \mathbf{y}_j \label{eq:origmixup_labels}
\end{align}
where $\alpha \in [0,1]$ is the mixup factor.
We generalize this principle by generating each training sequence as a combination of a random number of sources $I = [1,I_\text{max}]$.

Each source clip is augmented by various techniques to increase the \emph{source} diversity, such as random spectral and bandpass filtering, reverberation, random source levels with $\mathcal{N}(0,2)$~dB and pitch shifting and time stretching.
After adding multiple tracks together, we apply typical \emph{post-microphone} modifications such as audio codecs, non-linear distortion functions and varying signal levels.
\begin{algorithm}[tb]
	\caption{Modified mixup augmentation}\label{alg:mixup}
	\begin{algorithmic}
		\Require Sample class $c$, number of sources $I$, sequence length $L$
		\State $\mathbf{x} = \mathbf{0}$, $\mathbf{y} = \mathbf{0}$
		\For{$i<I$}
		\State $\mathbf{x}_i = []$, $\mathbf{y}_i = \mathbf{0}$
		\While{$len(\mathbf{x}_i)<L$}
		\State Sample $\mathbf{x}_\text{raw}, \mathbf{y}_\text{raw}$ with active sampled class $c$
		\State $\mathbf{x}_i = cat(\mathbf{x}_i, \mathbf{x}_\text{raw})$		
		\State $\mathbf{y}_i = \mathbf{y}_i \cup \mathbf{y}_\text{raw}$ \Comment{join sequence labels}
		\EndWhile
		\State $g_i \sim \mathcal{N}(0, \sigma_s)$ \Comment{Sample source gain}
		\State $\mathbf{x}_i = g_i \, f_\text{src}(\mathbf{x}_i)$ \Comment{source augmentation}
		\State $\mathbf{x} = \mathbf{x} + \mathbf{x}_i$ \Comment{add sources}
		\State $\mathbf{y} = \mathbf{y} \cup \mathbf{y}_i$ \Comment{join source labels}
		\EndFor
		\State $\mathbf{x} = f_\text{mic}(\mathbf{x})$ \Comment{mic augmentation}\\
		\Return $\mathbf{x}, \mathbf{y}$
	\end{algorithmic}
\end{algorithm}

\subsection{Label combination}
In vanilla mixup \eqref{eq:origmixup_data},\eqref{eq:origmixup_labels}, data and labels are mixed with the same linear factor $\alpha$. We argue this to be ill-posed targets for the multi-label detection problem at hand, and therefore propose to use a more data representative mixing, e.g., $\alpha_\text{data} \sim \mathcal{N}(0, \sigma)$ in dB and the labels are combined with an $OR$ operation
\begin{equation}
	y = y_1 \cup y_2, \dots 
\end{equation}
This results in true binary labels for mixed class training samples, and avoids fractional target probabilities, which matches the multi-label problem formulation.
%
%
The resulting data generation and augmentation procedure is outlined in Algorithm~\ref{alg:mixup}.

\subsection{Label enhancement}
\label{sec:labelcorrection}
Similarly as proposed in \cite{Fonseca2020,Gong2021}, we use a label correction or enhancement method relying on a teacher model. The teacher model is trained on the initially available labels, e.g.\ raw Audioset labels. Then, the teacher is used to obtain new predictions on the entire training data. We find that in many cases, the teacher models provides very confident probabilities for many sounds and often detects missing (\ac{FN}) or even wrongly set labels (\ac{FP}) in the Audioset labels. This motivates us to distinguish two steps: 
\begin{enumerate}
    \item When the teacher model contradicts the original label with high confidence, we correct this label. The corrected label $\tilde{y}_c$ is given when the prediction $\hat{y}_c$ contradicts the original label $y_c$ with very high confidence:
\begin{align}
    \tilde{y}_c &= 1 \quad\text{if}\quad y_c=0 \,\cap\, \hat{y}_c > 1-T \label{eq:FP}\\
    \tilde{y}_c &= 0 \quad\text{if}\quad y_c=1 \,\cap\, \hat{y}_c < T \label{eq:FN}
\end{align}
%
%
where $T$ is a low positive threshold of a few percent. The first equation \eqref{eq:FP} corrects \ac{FP} and \eqref{eq:FN} corrects \ac{FN}.
    \item When the teacher contradicts the original label with weaker confidence, this may be an ambiguous sample, difficult case or wrong label. Similar to \cite{Fonseca2020}, we propose to simply remove those labels from training, i.e., mask the labels out from participating in the loss. For less confident contradictions of the teacher model, the label is replaced with a flag ($NaN$), indicating that this label should be masked:
\begin{align}
	\tilde{y}_c &= NaN \quad\text{if}\quad (y_c=0) \,\cap\, (0.5 < \hat{y}_c < 1-T) \label{eq:mask_FP}\\
	\tilde{y}_c &= NaN \quad\text{if}\quad (y_c=1) \,\cap\, (T < \hat{y}_c < 0.5) \label{eq:mask_FN}
\end{align}
When computing the \ac{BCE} loss, $NaN$ labels are masked with a zero weight. 
\end{enumerate}

\subsection{CLAP as a baseline and zero-shot teacher}
The \ac{CLAP} model \cite{Elizalde2023,Elizalde2024} trains an audio-text encoder pair that maps paired audio-text data to close points in the embedding space. This trained encoder pair can then be used in a zero-shot manner for unseen tasks, such as audio classification, by computing the embedding similarity between an audio sample and various text prompts. It has been shown that CLAP has an astounding zero-shot performance for a wide variety of tasks, and can even outperform most state-of-the-art supervised models by supervised finetuning on the dedicated tasks.
We utilize CLAP both as a baseline and teacher model. As a proof of concept, we show that it is possible to generate training labels for ACDnet for an unlabeled audio dataset with CLAP for any sound class, and then use the proposed label correction to further improve the performance of ACDnet. As \ac{LLM} type text encoders are more accurate with more verbose prompts, we use the pre-fix "This is a sound of", as suggested in \cite{Elizalde2023}, with the class-specific prompts given in Tab.~\ref{tab:clap-prompts}. This gave us reasonably good results on the validation set, obviating the need for further prompt tuning.
\begin{table}[tb]
    \caption{Class short names and CLAP prompts}
    \label{tab:clap-prompts}
    \begin{tabular}{l|l}
        \toprule
        class 		& CLAP prompt \\\midrule
        voice 		& speech or singing \\
        music		& music \\
        cat			& cat meow \\
        dog 		& dog barking \\
        clapping 	& clapping or applause \\
        urban 		& an urban environment like traffic and city noise \\
        machinery 	& machines, tools and industrial sounds \\
        nature 		& nature sounds like wind, water, fire, animals \\
        windnoise 	& wind noise distorting the microphone \\
        alarm 		& alarms and sirens \\
        gunshot 	& gunshots and explosions \\
        \bottomrule
    \end{tabular}
\end{table}

Using CLAP to generate binary teacher labels is not straightforward for the multi-label case. Class-wise similarity scores are obtained by cosine similarity between the audio and per-class text embeddings. However, those scores can not be interpreted directly as probabilities. Typically, a softmax is applied, which implies a single active class at a time, i.e., monophony. However, as alternative normalization techniques resulted in worse performance for our multi-label validation and test sets, we use softmax normalization.

\section{Experiments}
\label{sec:experiments}

\subsection{Metrics and experimental parameters}
We evaluate with \ac{AP}, which measures the area under precision-recall curve, \cite{Davis2006} 
as threshold independent and data imbalance-agnostic evaluation metric. 
We show \ac{AP} per class and total average over all classes to obtain a balanced ranking.
The threshold for label enhancement is chosen with $T=0.05$.
We use a threshold of 0.2 to detect sound presence for CLAP with softmax, which was tuned on the validation set.

\subsection{Evaluation data}
To asssess the effect of training schemes using noisy training labels on real-world performance, we use FSD50k~\cite{fonseca2022FSD50K} as validation and test set, which have high quality polyphonic labels, so we can assume no label noise in the test sets. We merge the training and validation sets of FSD50k~\cite{fonseca2022FSD50K} as our validation set, and use the test set files for testing. To best match the receptive field of our network, we select files with a minimum length of 8 or 5 seconds, for validation and test sets, respectively. We sample 200 and 140 files, respectively, per active class to ensure a balanced evaluation. These choices were determined to strike a balance between having enough files per class and the longest possible minimum file length. 


\section{Results}
\label{sec:results}

\subsection{Proposed mixup and label correction}
\begin{figure}[tb]
	\includegraphics[width=1\columnwidth,clip,trim=0 10 0 30]{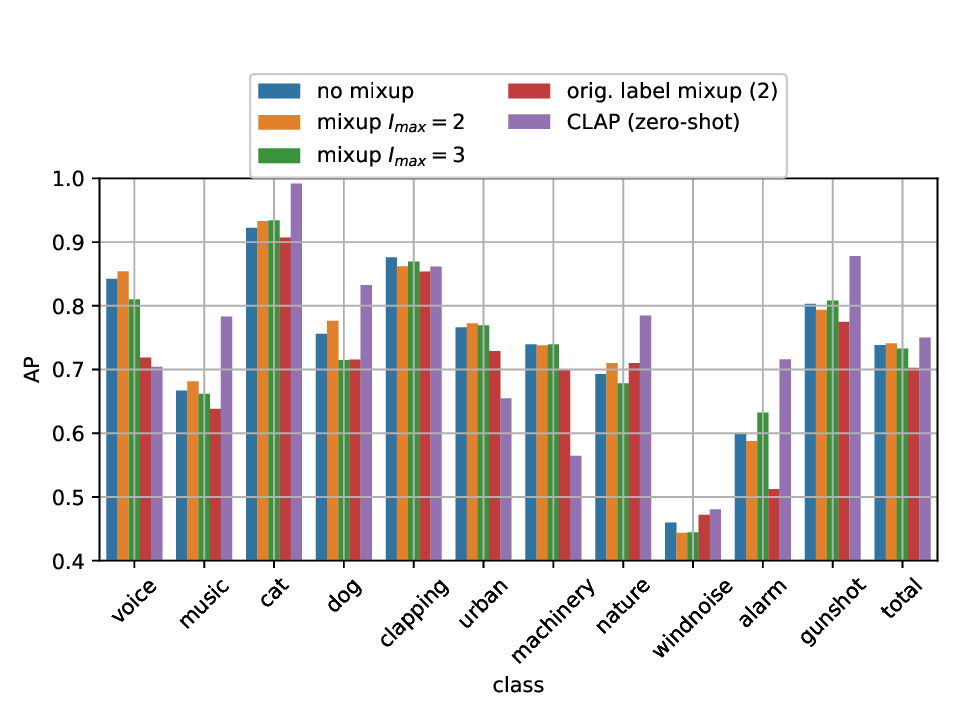}
	\caption{Test results for training with different mixup methods.}
	\label{fig:results_mixup}
\end{figure}

Fig.~\ref{fig:results_mixup} shows the \ac{AP} metric on the test set curated from FSD50k for no mixup, the proposed mixup with label joining with maximum mixup clips $I_\text{max} = \{2,3\}$, the original mixup interpolating the labels (red) with \eqref{eq:origmixup_labels}, and the zero-shot CLAP (version 2023) \cite{Elizalde2024} model as a baseline. 
Looking at the overall \ac{AP} averaged across all classes, the proposed mixup with $I_\text{max} = 2$ slightly outperforms no mixup, but $I_\text{max} = 3$ decreases \ac{AP} again. The differences are however surprisingly small. The proposed multi-label mixup with label joining shows improvements over the original mixup with weighted mixing of the labels via \eqref{eq:origmixup_labels}. This original mixup even shows a small detrimental overall effect compared to no mixup here. A reason for the low contribution of proposed multi-label mixup and detrimental original mixup may be that its effect becomes negligible when training on already polyphonic data, the large dataset size, and increased variety through augmentation.


\begin{table}[tb]
    \caption{Prior and proposed label correction strategies.}
	\label{tab:results_relabeling}
	\begin{tabular}{l|ccc}
		\toprule
		labels 		& Accuracy & mAP \\\midrule
        raw labels	                        & 0.899	& 0.741 \\\midrule
        percentile label masking (Fonseca2020)	& 0.897	& 0.750 \\
        percentile label corr (Fonseca2020*)	& 0.864	& 0.720 \\
        meanthresh label corr (Gong2021)	   & 0.892	& 0.754 \\\midrule
        label corr FN	                   & 0.902	& 0.751 \\
        label corr FN+FP	               & 0.900	& 0.756 \\
        label corr+masking FN	           & 0.899	& \textbf{0.767} \\
        label corr+masking FN+FP	       & \textbf{0.905}	& 0.762 \\
		\bottomrule
	\end{tabular}\\\vspace{4pt}
    {* modification of the baseline using correction instead of masking}
\end{table}

%
As a second experiment, we train ACDnet with the proposed mixup $I_\text{max} = 2$ on original and enhanced labels for Audioset to assess the contribution of the label correction and masking described in Sec.~\ref{sec:labelcorrection}. 
The class-average Accuracy and \ac{AP} (known as mAP) are shown in  Table~\ref{tab:results_relabeling}. We verify that for the baseline method~\cite{Fonseca2020} and their proposed missing label detection (percentile), correcting the labels is detrimental, while masking out wrong labels from training improves results. The second baseline \cite{Gong2021} using the average prediction over true positives as class-wise threshold to detect missing labels shows improvement over the raw noisy labels and the percentile label masking baseline.
We evaluate our proposed label enhancement step-by-step: correcting high confidence FN (missing labels) or both FN and FP yield improvements. When, in addition to the label correction,  labels deemed FN by the teacher with low confidence are masked out,
we observe an additional \ac{AP} boost. Correcting and masking both FN and FP does not yield a further improvement, and the difference between correcting  FN vs.\ FN+FP is minor. Therefore, the results for correcting FP are somewhat inconclusive, possibly due to the fact that the percentage of FPs in Audioset is much lower than FN.


\subsection{Training on unlabeled data}
\begin{figure}[tb]
	\includegraphics[width=1\columnwidth,clip,trim=0 10 0 30]{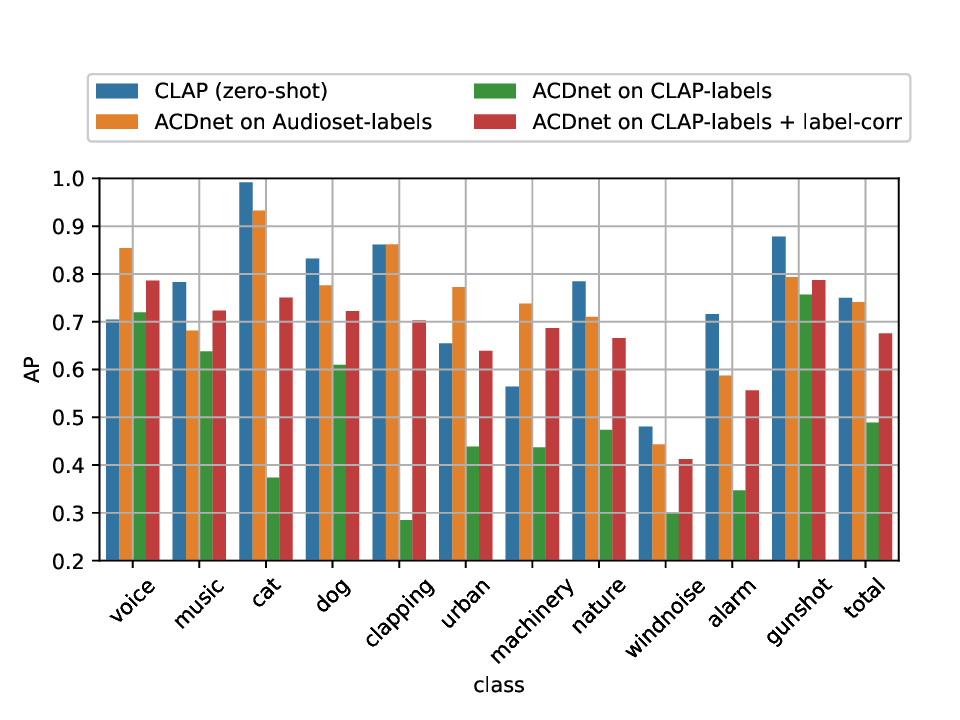}
    \caption{Training on unlabeled data: ACDnet trained on original Audioset labels as reference (orange), and ACDnet trained on Audioset with noisy teacher labels and with (red) or without (green) proposed label correction.}
	\label{fig:results_claplabels}
\end{figure}

Finally, as a proof of concept, we evaluate the feasibility of using an unlabeled dataset and generating noisy labels with the zero-shot CLAP model as a teacher. This is interesting as it can potentially be used to detect sound classes not available in labeled datasets.
Fig.~\ref{fig:results_claplabels} shows as reference again the zero-shot CLAP predictions on the test set (blue), and as second reference ACDnet trained on raw Audioset labels (orange). While training ACDnet on labels generated from CLAP (green) is feasible, it performs significantly worse compared to the two baselines. However, the performance can be greatly improved by using the proposed label correction (red), which increases the performance closer to the baseline trained on original Audioset labels (orange) and the initial teacher CLAP (blue). Note that ACDnet trained this way on unlabeled data (red) even outperforms CLAP for some classes while being computationally significantly more efficient and with a model size over 2 orders of magnitude smaller.


\section{Conclusion}
\label{sec:conclusion}
We proposed a multi-label model and training scheme for prevalent polyphonic sound detection tasks from noisy labels. The proposed model has a small compute footprint while achieving state-of-the-art detection precision on a real recorded test dataset with high quality labels. A new method to correct missing or wrong labels and mask out unreliable labels from training is shown to further improve the classification performance. Finally, we show that the proposed model can be trained on unlabeled data using teacher labels from a much larger zero-shot model, and that the proposed label enhancements can further improve model performance in this case.    

\newpage
\balance

\bibliographystyle{IEEEbib}
\bibliography{sapref.bib,refs.bib}

\end{document}